\documentstyle[12pt,epsfig]{article}

\begin{document}
\centerline{\large \bf A picture of the Yang-Mills deconfinement
transition}
\centerline{\large \bf and its lattice verification\footnote{Invited talk
presented by M.Engelhardt at the Eleventh International
Light-Cone Workshop on ``New directions in Quantum Chromodynamics'',
Kyungju, Korea, 21.-25.6.99, to appear in the proceedings.} }

\bigskip
\centerline{M.~Engelhardt\footnote{Supported by Deutsche 
Forschungsgemeinschaft under DFG En 415/1-1.},
K.~Langfeld, H.~Reinhardt\footnote{Supported by Deutsche
Forschungsgemeinschaft under DFG Re 856/4-1.}
and O.~Tennert}
\vspace{0.2 true cm}

\centerline{\em Institut f\"ur theoretische Physik, 
Universit\"at T\"ubingen}
\centerline{\em Auf der Morgenstelle 14, 72076 T\"ubingen, Germany}

\bigskip

\begin{abstract}
In the framework of the center vortex picture of confinement, the nature
of the deconfining phase transition is studied. Using recently developed
techniques which allow to associate a center vortex configuration with any
given lattice gauge configuration, it is demonstrated that the confining
phase is a phase in which vortices percolate, whereas the deconfined phase
is a phase in which vortices cease to percolate if one considers an
appropriate slice of space-time.
\end{abstract}

\section*{Heuristics of the center vortex picture}
A discussion of the deconfinement transition in Yang-Mills theory
presupposes a picture of the phenomenon of confinement. Conversely,
any picture of confinement should be able to accomodate the
deconfinement phase transition. The work presented here is concerned 
specifically with the so-called center vortex picture of confinement;
this picture is based on the conjectured presence of center vortices
in typical Yang-Mills gauge configurations. These vortices represent
closed magnetic flux lines in three space dimensions, describing 
closed two-dimensional world-sheets in four space-time
dimensions. Space-time in the following will always be considered
Euclidean. The magnetic flux represented by the vortices is furthermore
quantized such that a Wilson loop linking vortex flux takes a value
corresponding to a nontrivial center element of the gauge group. In the 
case of $SU(2)$ color discussed here, the only such element is $(-1)$.
For $N$ colors, there are $N-1$ different possible vortex fluxes
corresponding to the $N-1$ nontrivial center elements of $SU(N)$.

\begin{figure}[b!]
\vspace{-1.8cm}
\centerline{
\hspace{6cm}
\epsfxsize=4.8cm
\epsffile{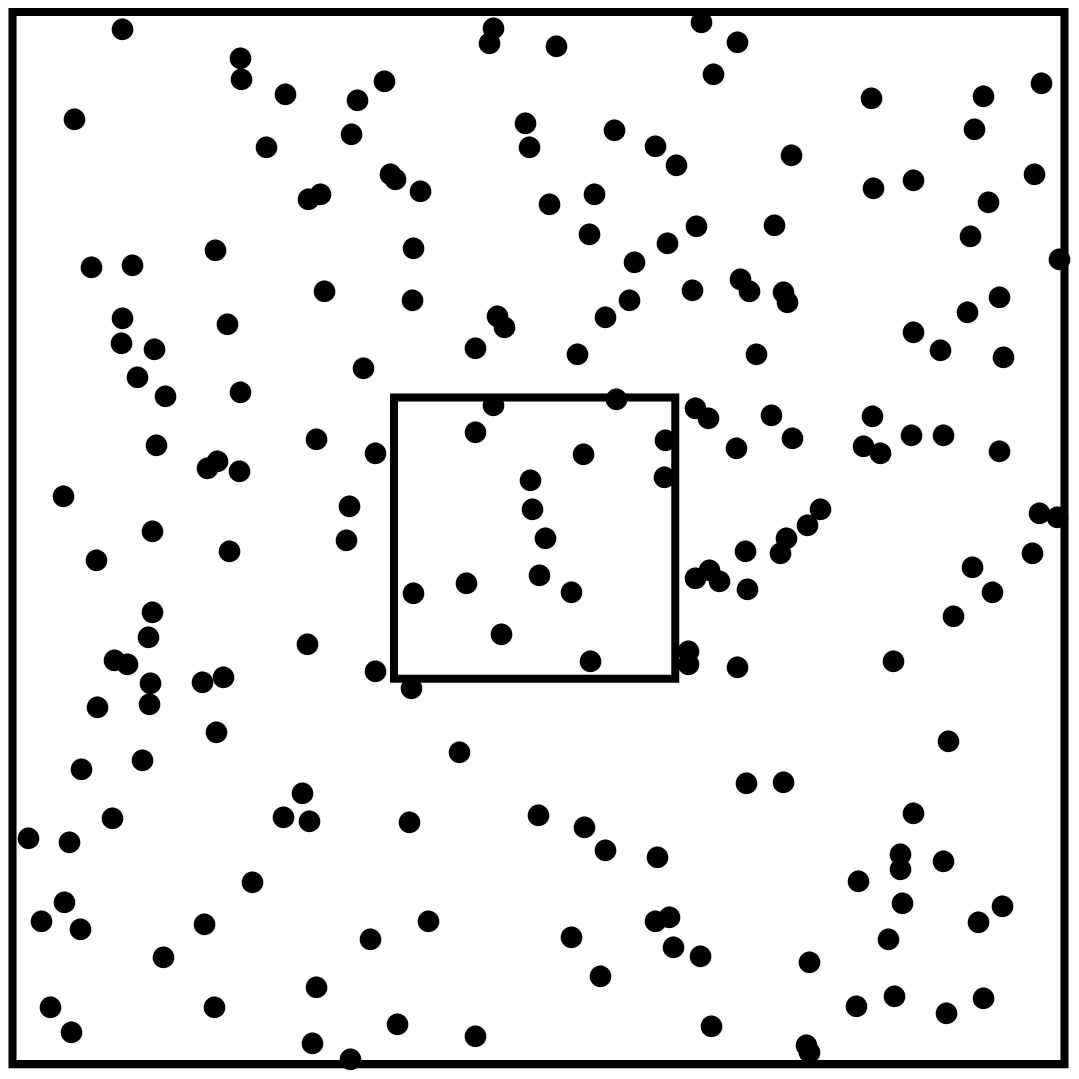}
\hspace{2cm}
\epsfxsize=4.8cm
\epsffile{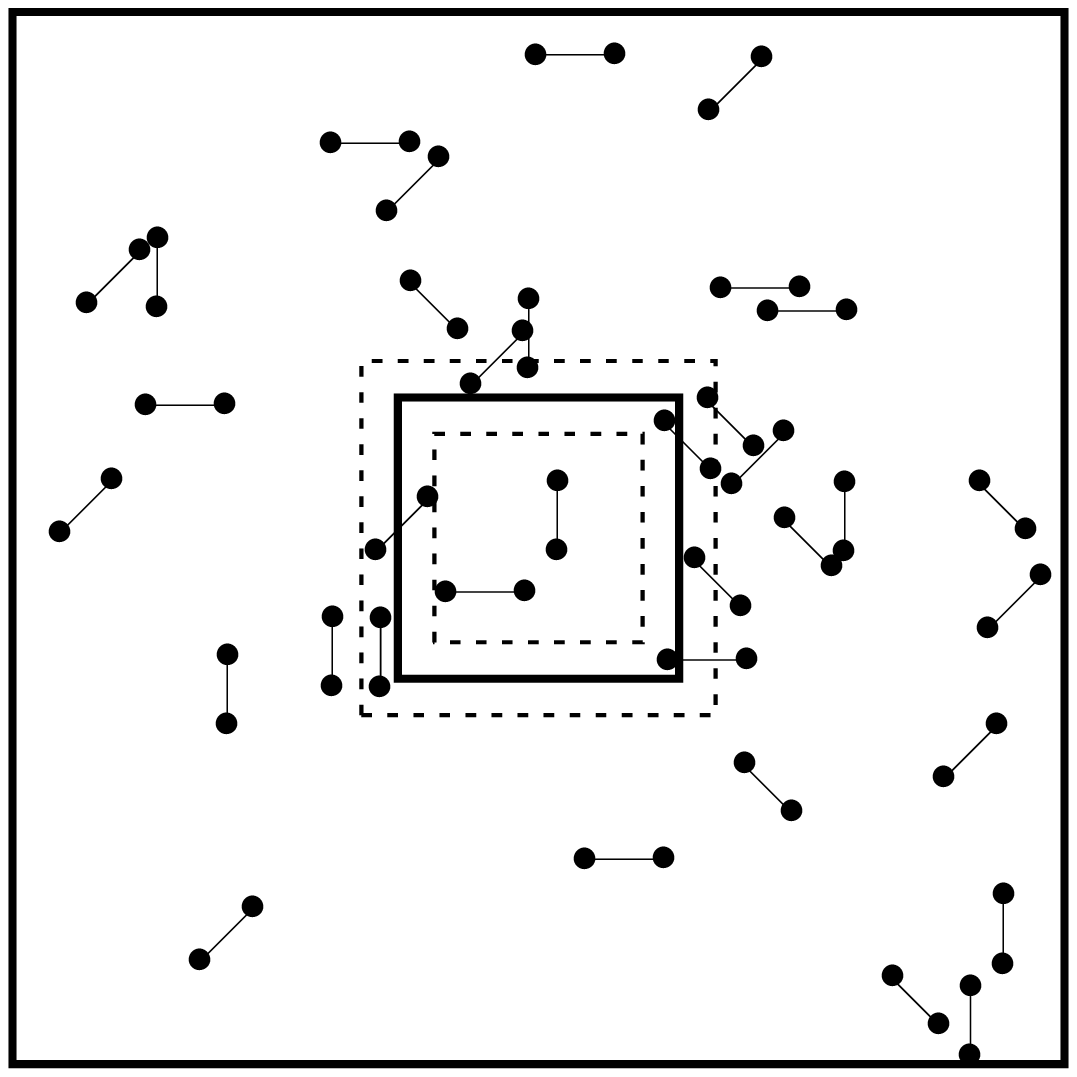}
}
\vspace{2.8cm}
\caption{Simple models for confining (left) and deconfining (right)
vortex ensembles.}
\label{fig1}
\end{figure}

Consider an ensemble of center vortex configurations in which the 
vortices are distributed randomly, specifically such that intersection 
points of vortices with a given two-dimensional plane in space-time
are found at random, uncorrelated locations. In such an ensemble,
confinement results in a very simple manner. Let the universe be
a cube of length $L$, and consider a two-dimensional slice of this
universe of area $L^2 $, with a Wilson loop embedded into it,
circumscribing an area $A$. On this plane, distribute $N$ vortex 
intersection points at random, cf. Fig.~\ref{fig1} (left).
According to the specification above, each of these points 
contributes a factor $(-1)$ to the value of the Wilson loop if 
it falls within the area $A$ spanned by the loop; the probability
for this to occur for any given point is $A/L^2 $.

The expectation value of the Wilson loop is readily evaluated in this simple
model. The probability that $n$ of the $N$ vortex intersection points
fall within the area $A$ is binomial, and, since the Wilson loop takes the 
value $(-1)^n $ in the presence of $n$ intersection points within the 
area $A$, its expectation value is
\begin{equation}
\langle W \rangle = \sum_{n=0}^{N} (-1)^{n}
\left( \begin{array}{c} N \\ n \end{array} \right)
\left( \frac{A}{L^2 } \right)^{n}
\left( 1 - \frac{A}{L^2 } \right)^{N-n}
=\left( 1 - \frac{2\rho A}{N} \right)^{N}
\stackrel{N\rightarrow \infty }{\longrightarrow }
\exp (-2\rho A)
\end{equation}
where in the last step, the size of the universe $L$ has been sent to
infinity while leaving the planar density $\rho =N/L^2 $ of vortex 
intersection points constant. Thus, one obtains an area law for the 
Wilson loop, with the string tension $\sigma = 2\rho $.

This simple mechanism lies at the core of the center vortex picture
of confinement. After having been proposed already in 
\cite{thoo,corn79}, evidence that the Yang-Mills dynamics actually
favors the formation of magnetic flux tubes arose in the framework
of the Copenhagen vacuum \cite{spag}. Also lattice studies were
initiated with the aim to study vortices \cite{mack,tomold}. 
These studies in essence defined vortices via their effect on Wilson 
loops, as discussed above. While this definition has the advantage of 
being gauge invariant, it does not allow to easily localize vortices, 
i.e. associate a collection of vortex world-surfaces with any given 
lattice gauge configuration. 

The absence of techniques
allowing to carry out such an identification for a long time posed a 
considerable obstacle to the study of center vortex physics, especially 
the study of their global properties. These properties, however, constitute
a crucial aspect for many applications, as a closer examination of the
above heuristic picture shows. Namely, for vortex intersection points to
be distributed in a sufficiently random manner on a space-time plane to
induce an area law for the Wilson loop, the vortices must form networks
which percolate throughout space-time. To see this, consider the converse,
namely that vortices can be separated into clusters of bounded extension.
This implies that any vortex intersection point on a plane comes with
a partner a finite distance (smaller than the bound on the cluster
extension) away, because vortices are closed. For simplicity, assume
the pairs of intersection points to occur with a fixed mutual distance $d$,
and distribute $N$ pairs on a space-time plane containing a Wilson loop
of area $A$, cf. Fig.~\ref{fig1} (right), where the lines between the 
points in the figure are merely to guide the eye in identifying pairs 
of points. Now, the probability that any given pair contributes a factor
$(-1)$ to the Wilson loop is $pPd/L^2 $, where $P$ denotes the perimeter
of the loop, since only pairs whose midpoints lie within a strip of
width $d$ around the Wilson loop are able to contribute a factor $(-1)$,
and they do this with a probability $p$ related to the angular distribution
of the pairs. Note that $p$ is independent of the dimensions of the Wilson
loop. The probability that $n$ pairs contribute a factor $(-1)$ is again
binomial, in complete analogy to above, and one consequently obtains 
a perimeter law for the expectation value of the Wilson loop in the
limit of an infinite universe,
\begin{equation}
\langle W \rangle
=\left( 1 - \frac{2pPd}{L^2 } \right)^{N}
\stackrel{N\rightarrow \infty }{\longrightarrow }
\exp (-\rho pPd)
\end{equation}
where $\rho =2N/L^2 $ again denotes the density of points. Thus, in the
absence of percolation, confinement disappears. This leads to the
conjecture that the deconfinement phase transition in the vortex
picture may take the guise of a percolation transition. However, as
already indicated above, to test such global properties of vortices in
lattice experiments, new techniques are needed which allow to 
associate a vortex world-sheet configuration with any given lattice
gauge configuration. These techniques have only been furnished quite
recently, sparking renewed interest in the vortex picture.
The present work is one contribution to these efforts.

\section*{Locating vortices on the lattice}
The abovementioned techniques, introduced in \cite{deb97,deb97aug,giedt},
employ a two-step procedure familiar from the dual superconductor 
picture of confinement. First, one uses the gauge freedom to bring a 
given gauge configuration as close as possible to the collective degrees 
of freedom under consideration; in the case of the dual superconductor, 
that is the Abelian degrees of freedom, in particular, the monopoles. 
The second step consists of projecting onto these degrees of freedom, 
i.e. neglecting residual deviations away from, say, Abelian configurations
in the case of the dual superconductor. This second step clearly
constitutes a truncation of the theory.

This idea was adapted to the case of vortex degrees of freedom as 
follows \cite{deb97,deb97aug,giedt}. One fixes gauge configurations to 
the {\em maximal center gauge},
\begin{equation}
\mbox{max } \sum_{i} | \mbox{tr } U_i |^{2}
\end{equation}
where the $U_i $ are the link variables on a space-time lattice. This
procedure biases links towards elements of the center of the gauge group.
Next, one performs a truncation of the configurations, namely
{\em center projection},
\begin{equation}
U \longrightarrow \mbox{ sign tr } U
\end{equation}
i.e. one replaces each $SU(2)$ link variable by the center element closest 
to it in the group. Thus, one remains with a lattice of center elements. 
Such a lattice can be associated in the standard fashion with a vortex 
configuration. One examines all plaquettes on the lattice, and if a 
plaquette takes the value $(-1)$, a vortex is said to pierce that plaquette. 
Thus, vortices in the lattice formulation are defined on the dual lattice,
i.e. the lattice shifted by the vector $(a/2,a/2,a/2,a/2)$ w.r.t. the
original one, $a$ denoting the lattice spacing. One can easily
convince oneself that the vortices defined in this way have all the
properties postulated further above.

Having isolated vortices on the lattice, the first question to answer
is whether these degrees of freedom do indeed determine the physics
of confinement, i.e. whether they furnish the full string tension
found in exact calculations without any truncations. Without this
basis, more detailed considerations of vortex properties run the
risk of being academical. One carries out two lattice experiments,
both times using the full Yang-Mills action as a weight, but in one
experiment, one calculates the observable in question, such as the
Wilson loop, using the full configurations; in the other experiment,
one uses the center projected configurations. If the results agree,
the observable is said to display {\em center dominance}. Center dominance
for the string tension has indeed been verified in $SU(2)$ lattice
gauge theory both at zero temperature \cite{deb97,deb97aug,giedt} and at 
finite temperatures \cite{tempv,tlang}, including the so-called 
``spatial string tension'' all the way into the deconfined regime.
Furthermore, the vortex density obeys the proper scaling 
law as dictated by the renormalization group for physical quantities, 
cf. \cite{kurt} (note erratum in \cite{tempv}) and \cite{giedt}.

\section*{Vortex percolation properties}
Given techniques allowing to locate vortex world-sheets in space-time,
or vortex loops on three-dimensional slices thereof, it is possible to 
discriminate between different vortex clusters. In the following,
three-dimensional slices of space-time, where one of the space directions
is left away, will be considered, since this displays the relevant 
percolation properties most clearly. To define a cluster, one finds a link 
on the dual lattice which is part of a vortex and furthermore locates all 
adjacent links which are also part of the vortex. This is repeated with all 
new links found, until no further links exist which are connected with the 
cluster in question. Having detected all vortex clusters in this manner, it 
is possible to determine the space-time extension of each cluster, i.e. the 
largest distance between any pair of points on the cluster. In a percolating
phase, most of the available vortex length will be organized into
clusters of the maximal possible extension, whereas in a phase with
no vortex percolation, most of the vortex material present in the
configuration will be concentrated in clusters much smaller than
the typical extension of the universe. 

\begin{figure}[b!]
\vspace{-0.8cm}
\centerline{
\hspace{3cm}
\epsfysize=8cm
\epsffile{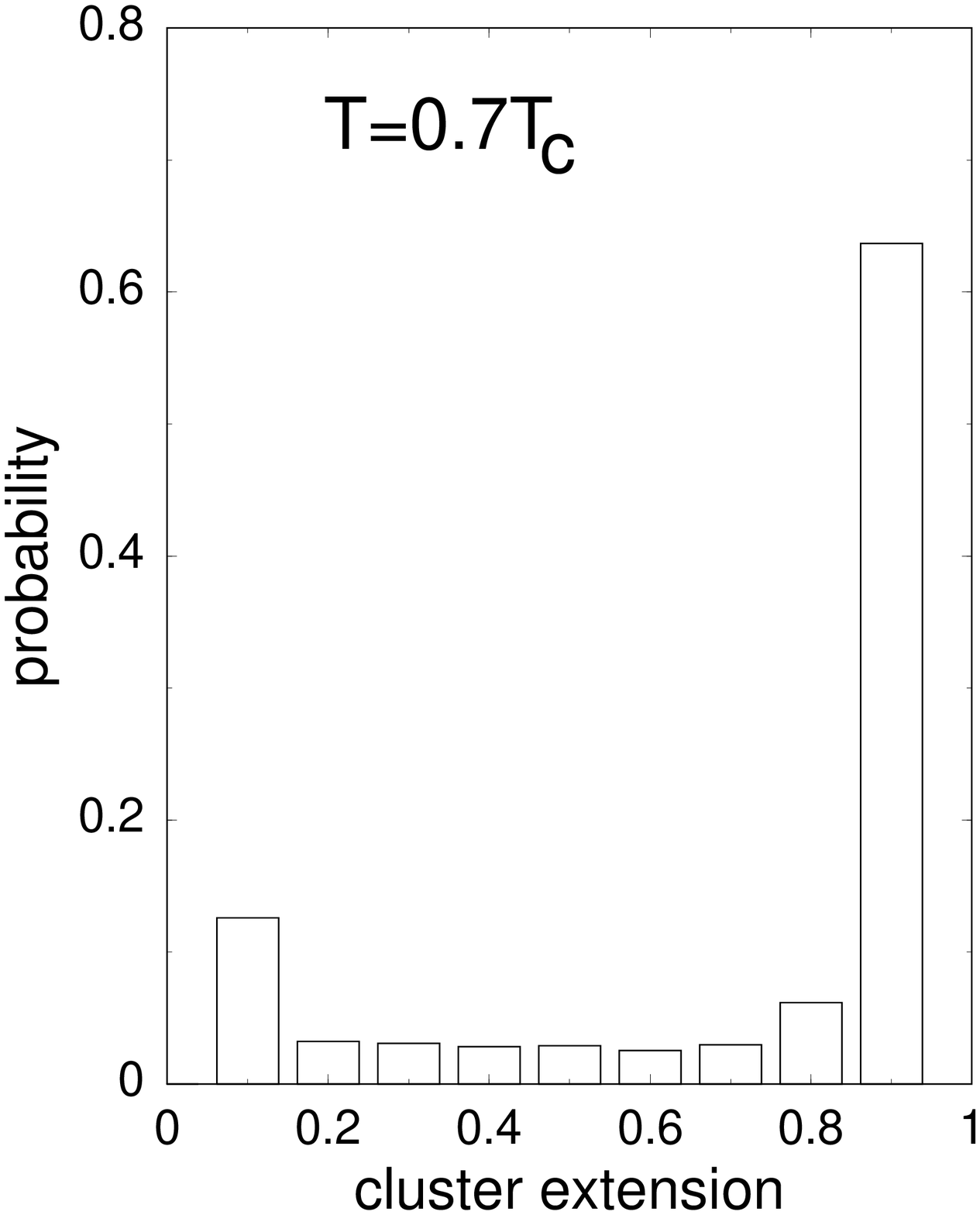}
\hspace{-2cm}
\epsfysize=8cm
\epsffile{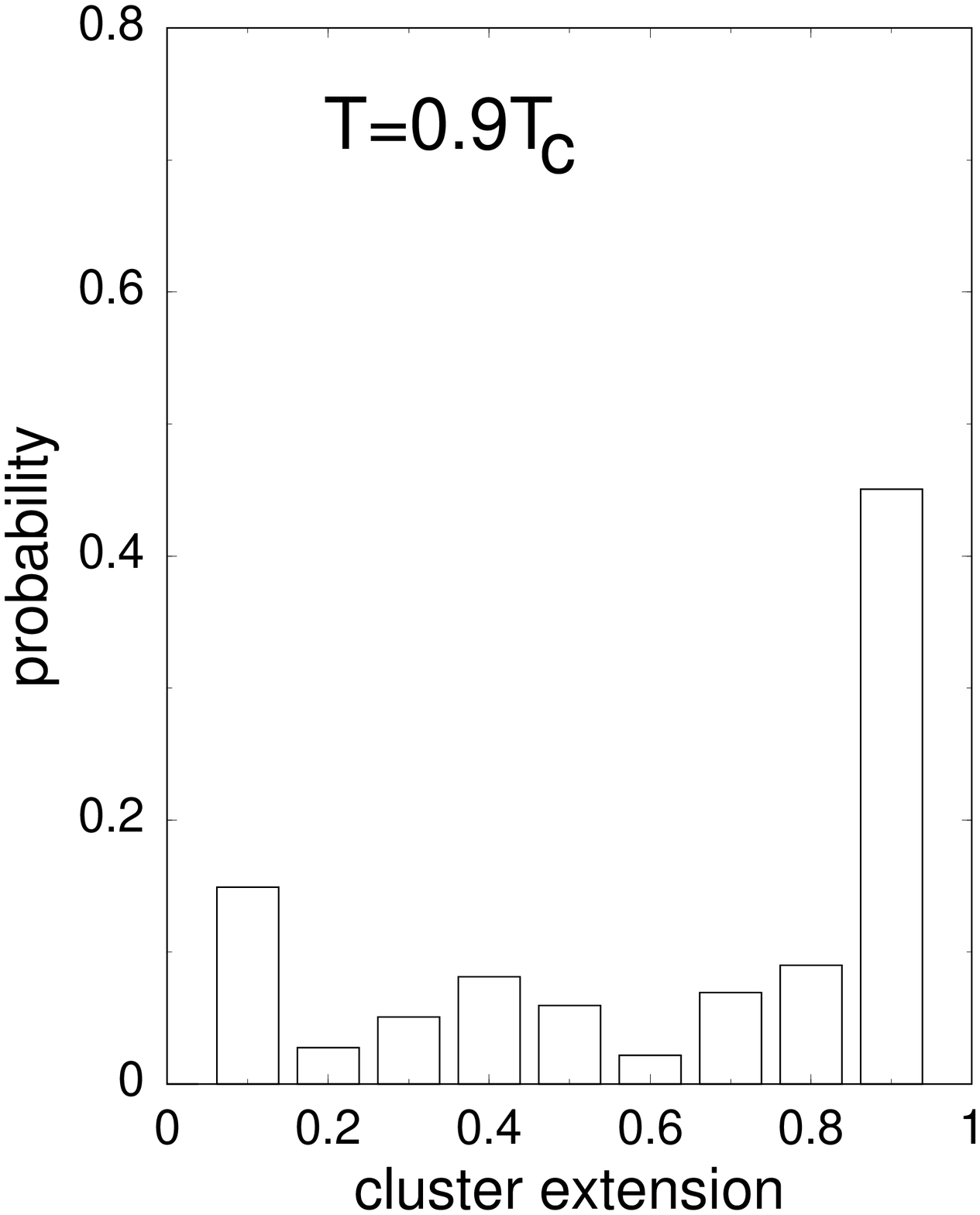}
}
\vspace{-0.2cm}
\centerline{
\hspace{3cm}
\epsfysize=8cm
\epsffile{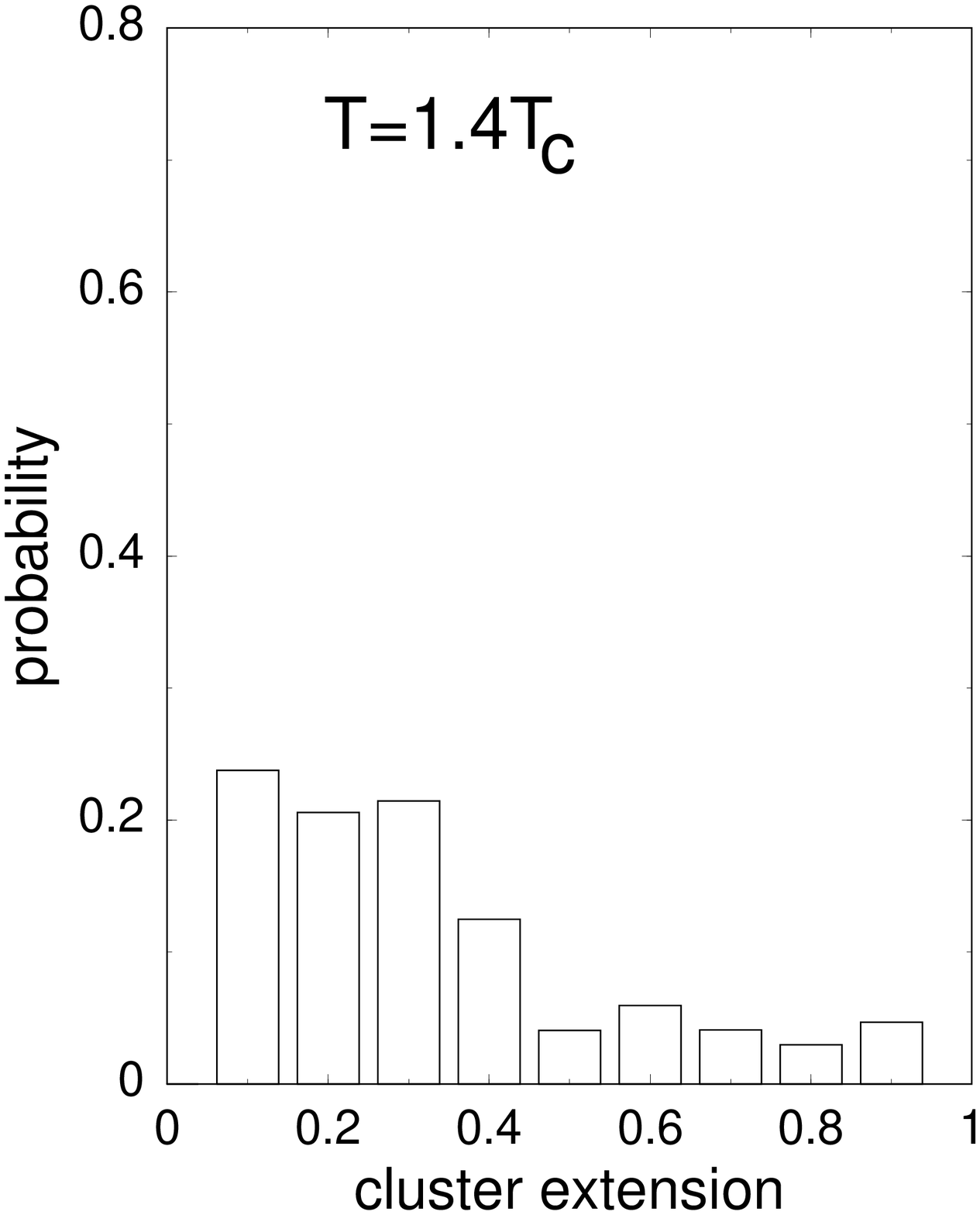}
\hspace{-2cm}
\epsfysize=8cm
\epsffile{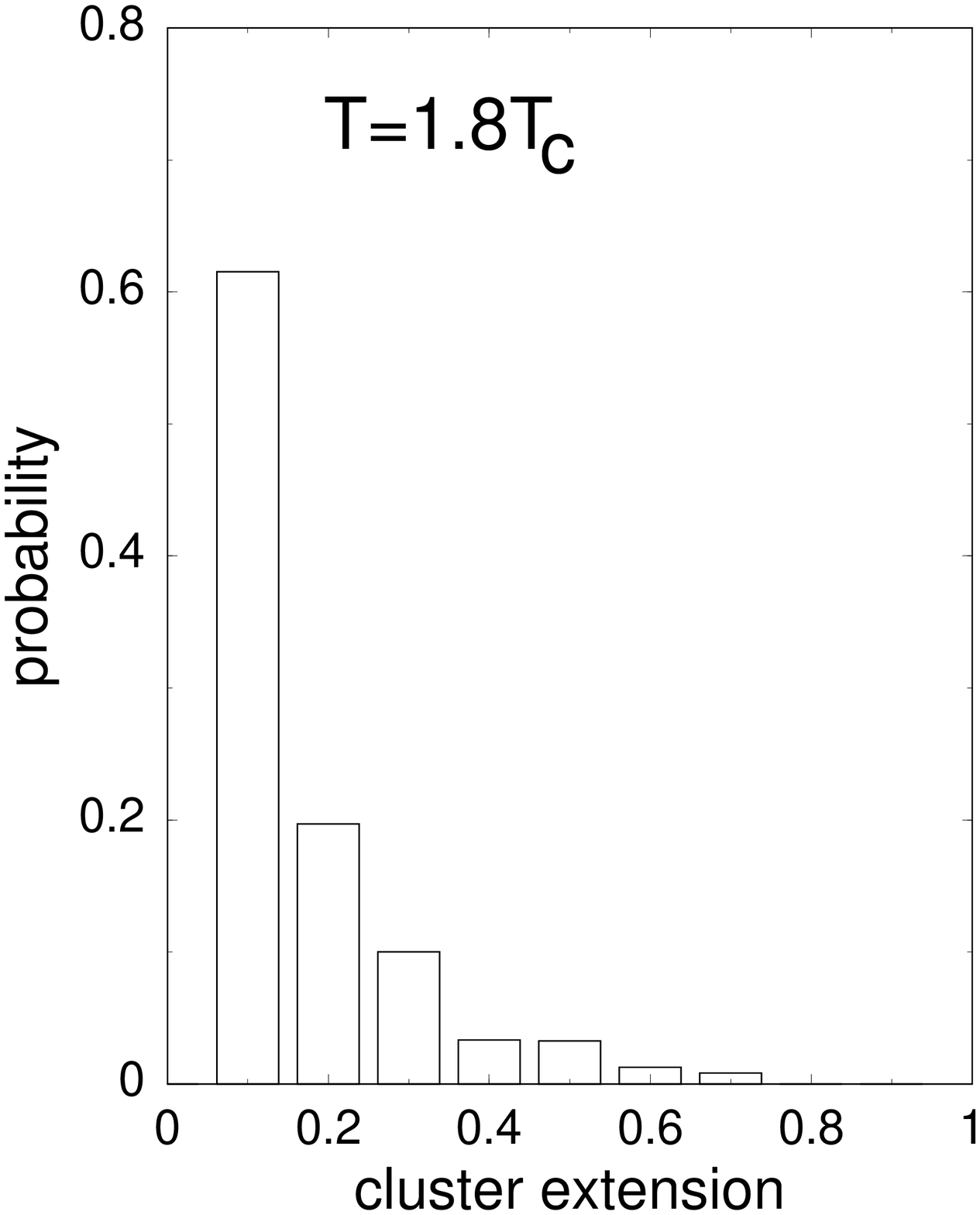}
}
\caption{Vortex material distributions.}
\vspace{-0.3cm}
\label{fig3}
\end{figure}

\begin{figure}[t!]
\vspace{-0.6cm}
\centerline{
\epsfxsize=6cm
\epsffile{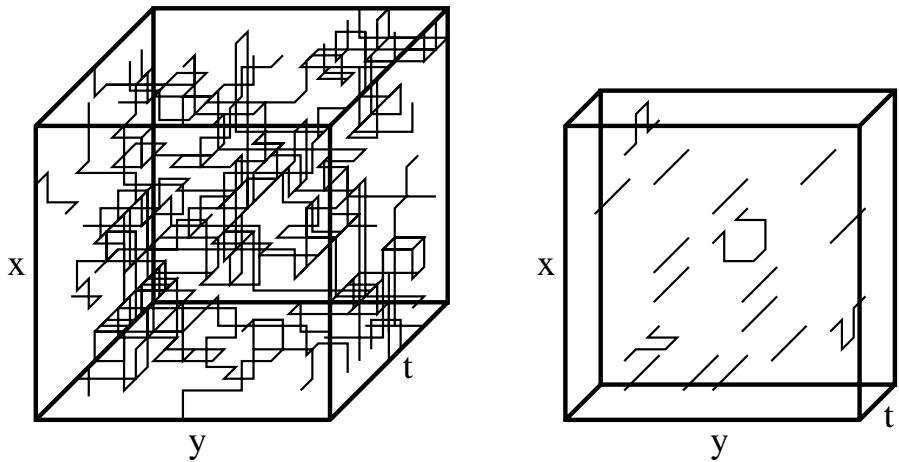}
}
\vspace{-0.4cm}
\caption{Typical vortex configurations in the confining (left) and the
deconfined phase (right).}
\label{fig4}
\end{figure}

To generate ``vortex material distributions'' which allow to read off 
which scenario is realized, one simply measures both the extension of 
each cluster as well as the number of links contained in it, and adds the 
latter number to a bin corresponding to the cluster extension in question. 
Fig.~\ref{fig3} displays such distributions, obtained for $\beta =2.4$ 
on $12^3 \times N_t $ lattices \cite{tlang}, which have been normalized 
such that the integral over the distributions gives unity, and where the 
cluster extension on the horizontal axis is in units of the maximal 
extension possible in the universe in question.
In view of Fig.~\ref{fig3}, one indeed obtains a 
transition from a confining phase, in which vortices percolate, to a 
deconfining phase, in which they cease to percolate. This confirms the 
conjecture proposed above in the introductory section. If one analyzes 
the small vortex clusters dominating the deconfined phase in more detail, 
one finds that a large part of these vortices wind in the (Euclidean) 
temporal direction, i.e. the space-time direction whose extension is
identified with the inverse temperature. Therefore, one finds that the 
typical configurations in the two phases can be characterized as displayed
in Fig.~\ref{fig4} in a three-dimensional slice of space-time, where one 
space direction has been left away. Note that Fig.~\ref{fig4}
also furnishes an explanation of the spatial string tension in the
deconfined phase. A spatial Wilson loop embedded into Fig.~\ref{fig4} (right)
can exhibit an area law, since intersection points of winding vortices
with the minimal area spanned by the loop can occur in an uncorrelated 
fashion despite those vortices having small extension. Note also the dual 
nature of this (magnetic) picture as compared with electric flux models 
\cite{patel}. In such models, electric flux percolates in the 
{\em deconfined} phase, while it does not percolate in the confining phase.

\section*{Outlook}
While it has thus been established {\em how} vortices generate the
confining and deconfining phases of Yang-Mills theory, it remains to
be clarified what the essential features of the {\em dynamics} underlying 
their behavior are. One interesting observation in this context is that a
simple model of vortices as random surfaces in four-dimensional
space-time already is able to generate the vortex phenomenology 
described above, i.e. a percolating confining and a non-percolating 
deconfining phase, separated by a transition as a function of temperature.
The necessary ingredients are an action per unit vortex area (i.e. a 
Nambu-Goto term), and an action penalty related to the curvature of the
vortex surfaces. By construction, this model can be understood in terms
of the entropy associated with random surfaces in a given space-time
domain; it contains no further dynamics. Evaluating the partition
function of such a model amounts to counting possible vortex surface 
configurations given a certain vortex density (enforced by the
Nambu-Goto term), and given an ultraviolet cutoff on the space-time
fluctuations of the surfaces (enforced by the curvature penalty).
A detailed report on a lattice investigation of this model will
be given in an upcoming publication.

Further issues being, or recently having been, investigated include:
The Pontryagin index associated with center vortex configurations
\cite{forcrand,cont}, and the breaking of
chiral symmetry \cite{forcrand}; the continuum meaning of the maximal
center gauge \cite{cont}; generalizations to $SU(3)$ color
\cite{giedt,montero}; and whether a random surface model for
vortices can be justified in terms of a low-energy effective theory
describing infrared Yang-Mills dynamics \cite{cont}.

\end{document}